\documentclass{elsart}
\usepackage{graphicx}
\usepackage{color}
\usepackage{amssymb}
\begin{document}
\begin{frontmatter}
\title{\mbox{\boldmath $\phi$}
photo-production from Li, C, Al, and Cu nuclei
at \mbox{\boldmath $E_\gamma=$}1.5--2.4~GeV}
\author[kyoto]{T.~Ishikawa\thanksref{now-lns}},
\ead{ishikawa@lns.tohoku.ac.jp}
\author[pusan]{D.S.~Ahn\thanksref{now-rcnp}},
\author[pusan]{J.K.~Ahn},
\author[konan]{H.~Akimune},
\author[taiwan]{W.C.~Chang},
\author[jasri]{S.~Dat\'{e}},
\author[rcnp]{H.~Fujimura\thanksref{now-kyoto}}, 
\author[rcnp,jaeri]{M.~Fujiwara},
\author[ohio]{K.~Hicks},
\author[rcnp]{T.~Hotta}, 
\author[kyoto]{K.~Imai},
\author[chiba]{H.~Kawai},
\author[rcnp]{K.~Kino},
\author[rcnp]{H.~Kohri},
\author[rcnp,jaeri]{T.~Matsumura\thanksref{now-nda}},
\author[rcnp,jaeri]{T.~Mibe\thanksref{now-jlab}},
\author[kyoto]{K.~Miwa},
\author[kyoto]{M.~Miyabe},
\author[rcnp]{M.~Morita},
\author[kyoto]{T.~Murakami},
\author[jaeri]{N.~Muramatsu\thanksref{now-rcnp}},
\author[osaka]{H.~Nakamura},
\author[kyoto]{M.~Nakamura\thanksref{now-wakayama}},
\author[rcnp]{T.~Nakano},
\author[kyoto]{M.~Niiyama},
\author[osaka]{M.~Nomachi},
\author[jasri]{Y.~Ohashi},
\author[chiba]{T.~Ooba},
\author[taiwan]{D.S.~Oshuev},
\author[canada]{C.~Rangacharyulu},
\author[osaka]{A.~Sakaguchi},
\author[chiba]{Y.~Shiino},
\author[rcnp]{Y.~Sakemi},
\author[lns]{H.~Shimizu},
\author[osaka]{Y.~Sugaya},
\author[osaka,jaeri]{M.~Sumihama\thanksref{now-tohoku}},
\author[miyazaki]{Y.~Toi},
\author[jasri]{H.~Toyokawa},
\author[taiwan]{C.W.~Wang},
\author[rcnp]{T.~Yorita\thanksref{now-jasri}},
\author[kyoto]{M.~Yosoi},
\author[rcnp]{R.G.T.~Zegers\thanksref{now-msu}}

\address[kyoto]{
\it Department of Physics, Kyoto University, Kyoto~606-8502, Japan}
\address[pusan]{
\it Department of Physics, Pusan National University, Busan~609-735, Korea}
\address[konan]{
\it Department of Physics, Konan University, Kobe~658-8501, Japan}
\address[taiwan]{
\it Institute of Physics, Academia Sinica, Taipei~11529, Taiwan}
\address[jasri]{
\it Japan Synchrotron Radiation Research Institute,
Mikazuki~679-5198, Japan} 
\address[rcnp]{
\it Research Center for Nuclear Physics, Osaka University,
Ibaraki~567-0047, Japan} 
\address[jaeri]{
\it Advanced Science Research Center, Japan Atomic Energy Research Institute, 
 Tokai~319-1195, Japan} 
\address[ohio]{
\it Department of Physics and Astronomy, Ohio University,
Athens, Ohio~45701, USA} 
\address[chiba]{
\it Graduate School of Science and Technology, Chiba University,
Chiba~263-8522, Japan} 
\address[canada]{
\it Department of Physics and Engineering Physics , University of Saskatchewan,
Saskatoon, Saskatchewan~S7N5E2, Canada} 
\address[osaka]{
\it Department of Physics, Osaka University, Toyonaka~560-0043, Japan} 
\address[lns]{
\it Laboratory of Nuclear Science, Tohoku University,
Sendai~982-0826, Japan}
\address[miyazaki]{
\it Physical Engineering Group, Miyazaki University,
Miyazaki~889-2192, Japan}
\thanks[now-lns]{
Present address: Laboratory of Nuclear Science, 
Tohoku University, Sendai~982-0826, Japan}
\thanks[now-rcnp]{
Present address: Research Center for Nuclear Physics, Osaka University,
Ibaraki~567-0047, Japan}
\thanks[now-kyoto]{
Present address: Department of Physics, Kyoto University, Kyoto~606-8502, Japan}
\thanks[now-nda]{
Present address: Department of Applied Physics,
National Defense Academy, Yokosuka~239-8686, Japan}
\thanks[now-jlab]{
Present address: Department of Physics and Astronomy, Ohio University, Ohio 45701, USA}
\thanks[now-wakayama]{
Present address: Department of Physics, Liberal Arts and Sciences,
Wakayama Medical University, Wakayama~641-8509, Japan}
\thanks[now-tohoku]{
Present address: Department of Physics, Tohoku University, Sendai~980-8578, Japan}
\thanks[now-jasri]{
Present address: Japan Synchrotron Radiation Research Institute,
Mikazuki~679-5198, Japan} 
\thanks[now-msu]{
Present address: National Superconducting Cyclotron Laboratory,
Michigan State University, Michigan 48824, USA}

\begin{abstract}
The photo-production of $\phi$ mesons from Li, C, Al, and Cu
at forward angles has been measured 
at $E_\gamma=1.5$--$2.4$~GeV.
The number of events for incoherent $\phi$ photo-production
is found to have a target mass number dependence of 
$A^{0.72\pm 0.07}$
in the kinematical region of $|t|\le 0.6$~${\rm GeV}^2/c^2$.
The total cross section of the $\phi$-nucleon
interaction, $\sigma_{\phi N}$, 
has been estimated as $35^{+17}_{-11}$~mb using the $A$-dependence of the $\phi$ photo-production yield
and a Glauber-type multiple scattering theory.
This value is much larger than $\sigma_{\phi N}$ in free space,
suggesting that the $\phi$ properties might change in the nuclear medium.
\end{abstract}

\end{frontmatter}
The modification of vector mesons in nuclear matter is of
interest in hadron physics.
A broadening of the width and/or a decrease of the mass have been predicted
for the $\phi$ meson in the nuclear medium
because of partial restoration of chiral symmetry~\cite{PHIMD1} or
the meson-nucleon interaction in the nuclear medium~\cite{PHIMD2,PHIMD3,PHIMD4}.
The mass shift of the $\phi$ meson has been
experimentally studied in the $p$-$A$ reaction at the normal nuclear
density~\cite{ENY00}, and in high-energy heavy-ion collisions~\cite{AKI96}.
However, no clear evidence has been observed.

According to the OZI rule, the total $\phi$-$N$ cross section, $\sigma_{\phi N}$, should be small
since the $\phi$ meson consists of almost pure $s\bar{s}$.
If $\sigma_{\phi N}$ in the nuclear medium is the same as that in free space,
the incoherent $\phi$ photo-production cross section from a nucleus,~$\sigma^{\rm inc}_A$,
is approximately proportional to the target mass number, $A$, 
since almost all the produced $\phi$ mesons are expected to go
outside the nucleus without interacting with a nucleon.
If $\sigma_{\phi N}$ becomes larger in the nuclear medium,
some fraction of photo-produced $\phi$ mesons 
would interact with nucleons in the nucleus
and disappear via inelastic reactions.
In this case, the $A$-dependence sizeably deviates from $\sigma_A\propto A^1$.

Only one measurement of $\phi$ photo-production from various nuclei
is reported at $E_\gamma=6.4$--$9.0$~GeV~\cite{MCC71},
where coherent production is dominant.
The value of $\sigma_{\phi N}$ is not accurately determined
from the data of coherent production.
On the other hand, $\sigma_{\phi N}$ in free space 
is well determined to be $7.7$--$8.7$~mb 
from the $\phi$ photo-production cross section on the proton,
$\left.d\sigma/dt\right|_{t=0}$, at $E_\gamma=4.6$--$6.7$~GeV,
where the energy dependence of the $\gamma$-$\phi$ coupling is assumed to 
be constant 
on the basis of the vector meson dominance model~(VDM)~\cite{BEH75}.
A quark model~\cite{LIP66} gives a prediction of 
$13.0\pm 1.5$~mb for $\sigma_{\phi N}$~\cite{BEH75}.
This value is deduced from the total $\pi^\pm p$ and $K^+p$ cross sections 
obtained at the high energy limit.
The obtained and predicted values of $\sigma_{\phi N}$ in free space are
 much smaller 
than other meson-nucleon total cross sections
$\sigma_{\omega N}$, $\sigma_{\rho N}$, and $\sigma_{\eta N}$~($\sim 30$~mb)~\cite{ADEP0}.

The $\sigma_{\phi N}$ in the nuclear medium can be determined by using 
a Glauber-type multiple scattering theory for incoherent production~\cite{MAR68}.
The incoherent production cross section from nuclei, $d\sigma_A^{\rm inc}/dt$,
is described as a product of 
the $\phi$ photo-production cross section on the nucleon,  $d\sigma_N/dt$,
and the effective nucleon number, $A_{\rm eff}$, 
which is a function of the nucleon density distribution and $\sigma_{\phi N}$.
The main background from coherent $\phi$ photo-production is
suppressed near the threshold energy
because the momentum-transfer is high 
even at $0^\circ$ due to the heavy mass of the $\phi$ meson.
Thus, the $\sigma_{\phi N}$ in the nuclear medium is expected to be determined
less ambiguously 
near the threshold energy as compared with those at high energies.

The experiment was carried out using the Laser-Electron Photon facility
at SPring-8~(LEPS). 
Photons were produced by backward Compton scattering with an ultra-violet Ar laser
from 8~GeV electrons in the storage ring.
The recoil electrons were momentum analyzed by a bending magnet in the storage ring,
and were detected by a tagging counter placed at the exit of the bending magnet.
The experimental setup is described elsewhere~\cite{NAK01}.

The nuclear targets used in the experiment were Li, C, Al, and Cu
with thicknesses of 100~mm, 36~mm, 24~mm, and 3~mm, respectively.
All the targets used were natural.
The Li target block was placed in a target box filled with Ar gas.
The windows of the target box were sealed with 50~$\mu$m Aramid sheets.
To minimize the difference of the acceptances
among different target thicknesses and
to reduce a systematic error caused in the acceptance correction,
each of the other three targets was set by dividing into three pieces
with the same center of gravity and with the same standard deviation of the position 
along the photon beam direction as those of the Li target.
To avoid the systematic errors due to the
change of the beam conditions,
targets were exchanged every two hours.

\begin{figure}
\begin{center}
\includegraphics*[scale=0.75]{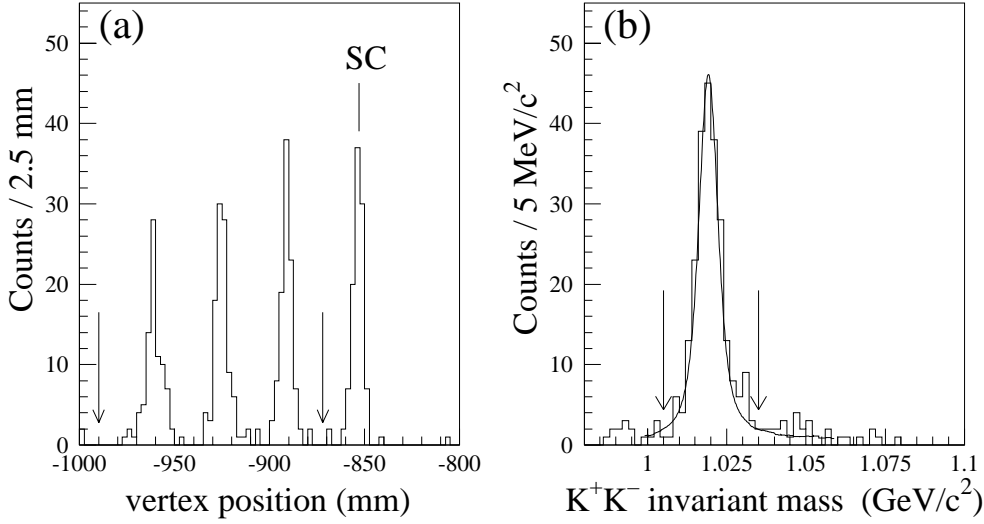}
\caption{
(a)~The vertex position distribution of the $K^+K^-$ events
along the photon beam direction for the $\gamma{\rm Cu}\rightarrow K^+K^-X$ reaction.
The cut points to select the $K^+K^-$ events generated at the target
are indicated by the arrows.
(b)~The $K^+K^-$ invariant mass spectrum for Cu.
The solid curve shows the fitting result with
the free $m_0$ and $\Gamma_0$ parameters given in the text.
The cut points to select the $\phi$ events are indicated by the arrows.
}
\label{fig:nuc}
\end{center}
\end{figure}

Charged particles produced at the target were detected at forward angles 
with the LEPS spectrometer which consisted of a dipole magnet,
a silicon-strip vertex detector, 
three multi-wire drift chambers,
a plastic scintillator behind the target~(SC),
and a plastic scintillator hodoscope placed
downstream of the tracking detectors~\cite{NAK01}.
A particle mass for each track was reconstructed by using
the time of flight and momentum information.
Kaons were identified within 4$\sigma$ of the 
mass resolution, which was momentum dependent and was
about 30~MeV/$c^2$ for 1 GeV/$c$ kaons.
The pion contamination in kaons due to particle mis-identification
was 3\% for 1 GeV/$c$ kaons.
The $\phi$ mesons produced in the targets were selected by utilizing 
the vertex position of the $K^+K^-$ events along the photon beam direction
as shown in Fig.~\ref{fig:nuc}~(a).
The position resolution at SC was typically 2~mm, and
the $K^+K^-$ events produced at SC was clearly separated.
Fig.~\ref{fig:nuc}~(b) shows the $K^+K^-$ invariant mass spectrum for the 
$\gamma{\rm Cu}\rightarrow K^+K^-X$ reaction.
A clear peak was observed,
and similar peaks were observed for the same reaction on other targets as well.

The measured mass and width are consistent 
with those of the free $\phi$ meson~\cite{PDG02}.
The experimental shape of the peak in the invariant mass spectrum
has been fitted by the sum of a
Breit-Wigner function convoluted with a Gaussian resolution function
and a background term,
\begin{equation}
N(m)=C \int_{-\infty}^{+\infty} L(m_0,\Gamma_0;m')
\frac{1}{\sqrt{2\pi\sigma_0^2}}e^{-(m-m')^2/2\sigma_0^2}
dm' + b B(m),
\end{equation}
where $\sigma_0$ denotes the resolution,
and $L(m_0,\Gamma_0;m)$ represents the Breit-Wigner function with the centroid of $m_0$
and the width of $\Gamma_0$.
The background term $B(m)$ is assumed to have a shape same as
those for non-resonant $K^+K^-$ production,
which are calculated by a Monte Carlo~(MC) simulation
with the assumption of 
the three-body phase space of the reaction $\gamma N\rightarrow K^+K^-N$.
In the case that $\sigma_0$ is fixed to the values predicted by the MC simulation,
the fitted mass $m_0$ and width $\Gamma_0$ of the $\phi$ meson 
are consistent with those in free space,
where the fitting region was 1000$-$1060~MeV$/c^2$.
The mass and width of the $\phi$ meson would
not change from its free-space values 
since
almost all the $\phi$ mesons decay outside a nucleus~($\gtrsim 95$\%)
in the momentum range from 1.0 to 2.2 GeV/$c$.
In the case that the $\sigma_0$ for each target is treated as a free parameter
instead of $\Gamma_0$,
the fitted value of $\sigma_0$ is  consistent
with the value estimated by the MC simulation.
The predicted $\sigma_0$ and the fitting results are summarized
in Table~\ref{tbl:fit}.
\begin{table}[htbp]
\begin{center}
\caption{
  The summary of the invariant mass resolution $\sigma_0$
  and the fitting results.
  (a)~The invariant mass resolutions predicted by the Monte Carlo~(MC) simulation.
  (b)~The fitting results 
     in the case that $\sigma_0$ is fixed to be 
     the value estimated in the MC simulation.
  (c)~The fitting results 
     in the case that the width of the $\phi$ meson, $\Gamma_0$, for each target is fixed
     to be the 
     same as that of the free $\phi$ meson ($4.26$~MeV/$c^2$~\cite{PDG02}).
}
\begin{tabular}{clrrrr}
\hline
\multicolumn{2}{c}{parameter~(MeV/$c^2$)} & 
\multicolumn{1}{c}{Li} &
\multicolumn{1}{c}{C}  &
\multicolumn{1}{c}{Al} & 
\multicolumn{1}{c}{Cu} \\
\hline
(a)&$\sigma_0$          & $1.6\pm 0.1$    & $1.9\pm 0.1$    & $2.3\pm 0.1$    & $2.1\pm 0.1$    \\
(b)&$m_0^{\rm fit}$     & $1019.7\pm 0.2$ & $1020.1\pm 0.3$ & $1019.5\pm 0.3$ & $1019.3\pm 0.3$ \\
&$\Gamma_0^{\rm fit}$   & $3.4\pm 0.4$    & $5.0\pm 0.7$    & $4.9\pm 0.8$    & $4.9\pm 0.8$    \\
(c)&$m_0^{\rm fit}$     & $1019.7\pm 0.2$ & $1020.1\pm 0.3$ & $1019.5\pm 0.3$ & $1019.3\pm 0.3$ \\
&$\sigma_0^{\rm fit}$   & $1.3\pm 0.4$    & $2.0\pm 0.5$    & $2.3\pm 0.4$    & $2.4\pm 0.4$    \\
\hline
\end{tabular}
\label{tbl:fit}
\end{center}
\end{table}

To determine the background subtracted $\phi$-yield normalized by 
incident photon numbers,
first, the $\phi$ meson events, $N_{KK}$, are selected by gating the $K^+K^-$ events
in the $K^+K^-$ invariant mass spectrum from 1005 to 
1035~MeV/$c^2$~(see Fig.1(b)).
The number of background events in the $\phi$ peak region, $N_{\rm BG}$,
are estimated by
\begin{equation}
 N_{\rm BG} = N_{\rm tail} \frac{N_{\rm peak}^{\rm MC}}{N_{\rm tail}^{\rm MC}},
\end{equation}
where $N_{\rm tail}$ means the observed number of events in the tail region
from 1050 to 1100~MeV/$c^2$.
The $N_{\rm peak}^{\rm MC}$ and $N_{\rm tail}^{\rm MC}$ are the estimated
number of events in the region of 1005--1035 and 1050--1100~MeV/$c^2$,
respectively, for the calculated non-resonant $K^+K^-$ events.
The fraction of the background events is small~(5--7\%).
The background events due to mis-identification of a pion as a kaon
are negligibly small.

The number of events for $\phi$ photo-production cross section 
is normalized by taking into account the number of hits in the tagging counter, $N_{\rm tag}$,
the attenuation of the photon flux in the target material, $\eta_{\rm att}$,
the number of target nuclei, $N_\tau$,
the live time of the data taking system, $\eta_{\rm DAQ}$,
and the acceptance correction, $\eta_{\rm acc}$.
The normalized number of events, $Y$, is then written as
\begin{equation}
Y= \frac{N_{KK}-N_{\rm BG}}{N_{\rm tag} N_\tau \eta_{\rm att} \eta_{\rm DAQ}\eta_{\rm acc}}.
\label{eq:yld}
\end{equation}
The normalized number of $\phi$ photo-production events
for each target is estimated by averaging data
for horizontally and vertically polarized incident photons to reduce the ambiguity
in acceptance correction due to the different decay asymmetry~\cite{MIB04}.
The transmission rate, i.e. survival rate of tagged photons at the target position,
is needed for determining the $\phi$ photo-production cross section.
However, $\sigma_{\phi N}$ can be determined directly from the normalized number of events
which is proportional to the $\phi$ photo-production cross section
since the transmission rate is the same for all the targets.
The number of events and the  normalization factors are summarized in Table~\ref{tbl:cor}.
\begin{table}[htbp]
\begin{center}
\caption{
Summary of the number of events and the normalization factors.
$N_{KK}$: the observed number of the $K^+K^-$ events in the $\phi$ peak region,
$N_{\rm BG}$: the estimated number of background events in the $\phi$ peak region,
$N_{\rm tag}$: the tagged photon flux, 
$\eta_{\rm att}$: the attenuation of the photon flux in the target material,
$N_\tau$: the number of target nuclei, 
$\eta_{\rm DAQ}$: the live time of the data taking system, and 
$\eta_{\rm acc}$: the acceptance correction.
}
\begin{tabular}{crrrr}
\hline
\multicolumn{1}{c}{} & 
\multicolumn{1}{c}{Li} &
\multicolumn{1}{c}{C}  &
\multicolumn{1}{c}{Al} & 
\multicolumn{1}{c}{Cu} \\
\hline
$N_{KK}$ &  348 & 267 & 286 & 238 \\
$N_{\rm BG}$    & 21.3$\pm$4.5 & 16.4$\pm$4.1 & 21.4$\pm$4.5 & 11.2$\pm$3.2 \\
$N_{\rm tag}\times 10^{-10}$ & 7.15 & 6.97 & 9.76 & 23.31 \\
$N_\tau\times 10^{-23}$& 4.63 & 3.12 & 1.45 & 0.254 \\
$\eta_{\rm att}$ & 0.976 & 0.948 & 0.906 & 0.926 \\  
$\eta_{\rm DAQ}$ & 0.922 & 0.899 & 0.874 & 0.911 \\
$\eta_{\rm acc}\times 10^2$ &
  $5.84\pm 0.06$ & $5.74\pm 0.06$ & $5.79\pm 0.07$ & $6.12\pm 0.07$ \\
\hline
\end{tabular}
\label{tbl:cor}
\end{center}
\end{table}

\begin{figure}
\begin{center}
\includegraphics*[scale=0.75]{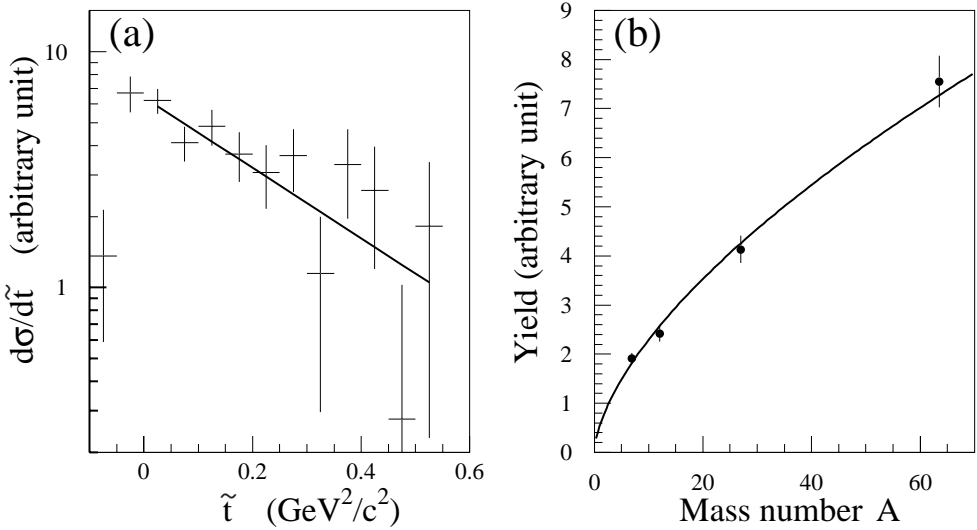}
\caption{
(a)~The $\tilde{t}=|t|-|t|_{\rm min}$ distribution for the
$\gamma{\rm C}\rightarrow K^+K^-X$ reaction.
The data points are fitted with the form $d\sigma/d\tilde{t}=C\exp(-b\tilde{t})$.
(b)~The $A$-dependence of the $\phi$ meson photo-production from nuclei.
The data points are fitted with the parameterization $A^{0.63}$.
}
\label{fig2}
\end{center}
\end{figure}
The measured momentum-transfer square $|t|$ ranges up to 0.6~GeV${}^2/c^2$.
Fig.~\ref{fig2}~(a) shows the $\tilde{t}=|t|-|t|_{\rm min}$ distribution
for C, where $|t|_{\rm min}$ is the minimum $|t|$ given under the assumption that
the target is a proton at rest.
The $\tilde{t}$ distribution is fitted with a function of 
$d\sigma/d\tilde{t} = C \exp (-b\tilde{t})$
in the region of $\tilde{t}=0.0$--$0.5$~GeV${}^2/c^2$.
The slope parameters $b$ are $3.6\pm 0.9$,
$4.5\pm 1.0$, $3.1\pm 0.9$, and $4.5\pm 1.0$~$({\rm GeV}^2/c^2)^{-1}$
for Li, C, Al, and Cu, respectively.
Any of these slope parameters is consistent with that for
$\phi$ photo-production on the proton,~$b=3.38\pm 0.23$~$({\rm GeV}^2/c^2)^{-1}$~\cite{MIB04},
within the errors.
The number of all the $\phi$ events for each target is
found to be proportional to $A^{0.63\pm 0.05}$
as shown in Fig.~\ref{fig2}~(b).

In order to determine $\sigma_{\phi N}$ from the $A$-dependence
of the $\phi$ photo-production yield,
an optical model of a Glauber-type multiple scattering theory
for incoherent production is applied~\cite{MAR68}.
In this model, the production cross section from a nucleus,
$d\sigma^{\rm inc}_A/dt$, is described as
\begin{equation}
\frac{d\sigma^{\rm inc}_A}{dt}=A_{\rm eff} \frac{d\sigma_N}{dt},
\label{eq:aeff0}
\end{equation}
where $A_{\rm eff}$ is the effective nucleon
number and $d\sigma_N/dt$ is the production cross section on the nucleon.
The $A_{\rm eff}$ for $\phi$ photo-production
is expressed as a function of $A$, $\sigma_{\gamma N}$,
and $\sigma_{\phi N}$;
\begin{equation}
\begin{array}{l}
A_{\rm eff}(A,\sigma_{\gamma N},\sigma_{\phi N})
=
\displaystyle 
\frac{1}{\sigma_{\phi N} - \sigma_{\gamma N}}
\int
\left(
e^{-\sigma_{\gamma N}T(\vec{b})}-e^{-\sigma_{\phi N}T(\vec{b})}
\right)\,
d^2b,\\
T(\vec{b})=A\int_{-\infty}^{+\infty} \rho(\vec{b},z)\, dz,
\end{array}
\label{eq:aeff}
\end{equation}
where $\sigma_{\gamma N}$ stands for the total photon-nucleon cross section, 
$\vec{b}$ denotes the impact vector of the incident photon, and $\rho$ is
the nucleon density of the target nucleus.
The effect of quasi-elastic collision between a $\phi$ meson and
a nucleon in the nucleus is not included in Eq.~(\ref{eq:aeff0}) since the direction and
energy change of the outgoing $\phi$ meson is small
because of the small direct $\phi NN$ coupling~\cite{CAB03}.
Assuming the same $d\sigma_N/dt$ for the proton and for the neutron,
$\sigma_{\phi N}$ can be derived from the $A$-dependence of the 
$\phi$ photo-production cross sections.
In this case,
the absolute values of $d\sigma_A/dt$ are not necessary.
The $\sigma_{\gamma N}$
is fixed to be 140~$\mu$b in the energy range from 1.5 to 2.4~GeV~\cite{PDG02}.
The nucleon density is
given by normalizing the charge density distribution~\cite{VRI87},
where the proton and neutron density distributions are assumed to have the same
$r$-dependence.
The same branching ratio of the $\phi\rightarrow K^+K^-$ process 
for each target nucleus is used
since almost all the $\phi$ mesons decay outside the nucleus.
The value of $\sigma_{\phi N}$ is estimated to be $71^{+32}_{-19}$~mb
from the $A$-dependence of the number of all the $\phi$ events.
This value is much larger than other meson-nucleon total cross sections.
This is attributed to the coherent production contribution as described below.

\begin{figure}
\begin{center}
\includegraphics*[scale=0.75]{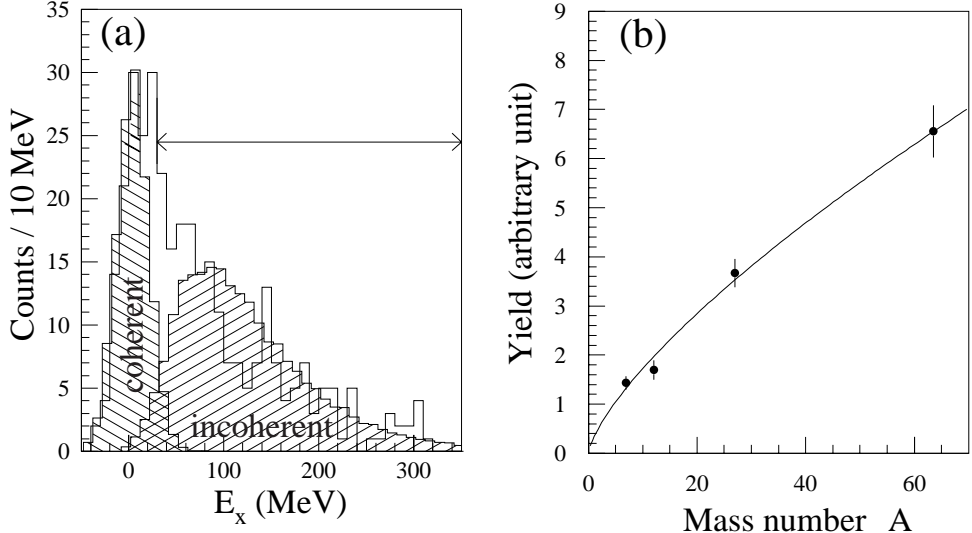}
\caption{
(a)~The missing energy spectrum of $\phi$ photo-production for the Li target.
The hatched regions show the calculated 
spectra for the coherent and incoherent processes
by the Monte Carlo simulation, and the normalization is made to guide the eyes.
(b)~The $A$-dependence of the number of events for $\phi$ photo-production
from nuclei after the contribution of coherent production is subtracted.
The data points are fitted with the parameterization $A^{0.72}$.
}
\label{fig3}
\end{center}
\end{figure}

It is reported in Ref.~\cite{CAB03} that the contribution of the coherent process
cannot be negligibly small especially for light nuclear targets
even at $E_\gamma\sim 2$~${\rm GeV}$.
The coherent production in Li has been evaluated in the missing
energy spectrum.
The missing energy, $E_x$, is defined as 
\begin{equation}
E_x = m_X-m_A,
\end{equation}
where $m_X$ is the missing mass in the reaction $\gamma A\rightarrow \phi X$,
and $m_A$ stands for the mass of the target nucleus.
Fig.~\ref{fig3}~(a) shows the missing energy spectrum for Li
together with the simulation results for the coherent $\gamma A\rightarrow \phi A$ and the incoherent 
$\gamma N\rightarrow \phi N$ processes.
The Fermi motion and the binding energy are taken into account for the incoherent
process in the MC simulation.
The missing energy spectrum of coherent $\phi$ photo-production
concentrates at 0~MeV within the experimental
resolution, and that of incoherent production is distributed in the positive energy region.
The missing energy spectrum for the coherent process 
is symmetrical around 0~MeV with a resolution of 19~MeV~($\sigma$).
Assuming that there are no $\phi$ events produced incoherently
in the negative $E_x$ region,
the total coherent events are estimated to be twice the number of
events in the negative $E_x$ region.
The number of the coherent $\phi$ photo-production events in Li is then 82.0$\pm$12.8.

Since the coherent contribution is relatively small for the heavier target,
the incoherent events in the negative $E_x$ region may not be negligible.
The coherent $\phi$ contributions in the other targets are evaluated theoretically
using the estimated one in Li as an input.
The contribution of the coherent process is proportional
to the square of the nuclear form factor~\cite{OSETP},
\begin{equation}
\frac{d\sigma}{dq}\propto|A\,F(q)|^2,
\end{equation}
where $q$ is the three dimensional momentum-transfer.
The coherent contribution is evaluated by integrating $d\sigma/dq$
over the kinematically allowed region of $q$.
The number of $\phi$ mesons produced coherently is 
then estimated to be 
72.9$\pm$11.4, 30.9$\pm$4.8, and
30.4$\pm$4.7 for C, Al, and Cu, respectively.
After subtracting the coherent contribution as the background,
the normalized number of events gives a relation $\sigma_A\propto A^{0.72\pm 0.07}$
as shown in Fig.~\ref{fig3}~(b),
from which $\sigma_{\phi N}$ is deduced, and is found to be $35^{+17}_{-11}$~mb.
As a cross check, the coherent contributions for the other targets
are estimated using the exactly same technique as in the case of Li.
In this case, $\sigma_A$ is proportional to $A^{0.73\pm 0.07}$, and 
$\sigma_{\phi N}$ is estimated to be $34^{+17}_{-11}$~mb.
These are consistent with the the former results.
Similar results are also obtained by selecting the kinematical
region for the incoherent process instead of subtracting the coherent
contribution. When the events with $E_x$ larger than 30~MeV are selected,
$\sigma_A\propto A^{0.74\pm 0.06}$
and $\sigma_{\phi N}=30^{+12}_{-8}$~mb are obtained.
The results are stable even if the missing energy cut is tightened up to 80~MeV.
These values obtained in this experiment
are much larger than $\sigma_{\phi N}$ in free space,
indicating the modification of the $\phi$-$N$ scattering amplitude
in the nuclear medium.

\begin{figure}
\begin{center}
\includegraphics*[scale=0.75]{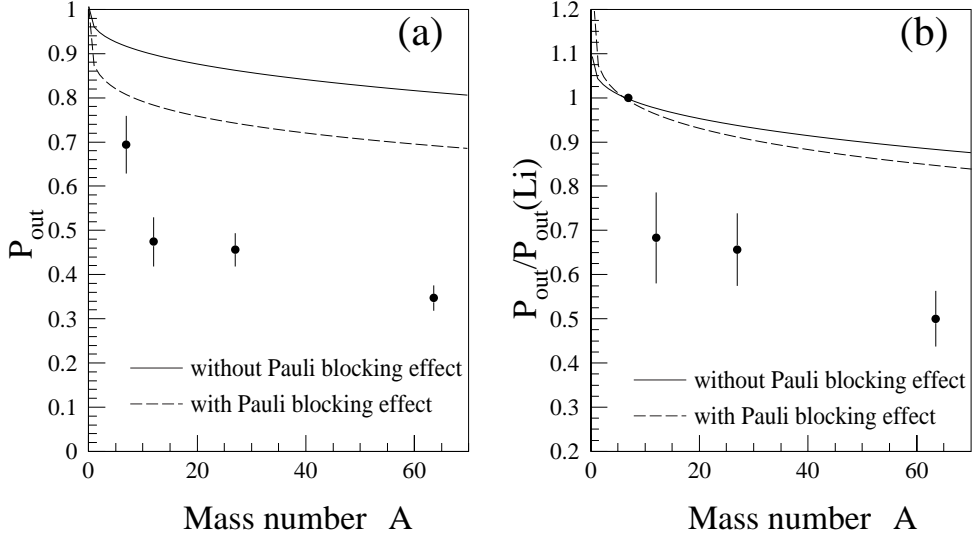}
\caption{
(a)~The probability $P_{\rm out}\!=\!\sigma_A/(A\sigma_N)$.
The overall normalization error~(18\%)
is not included. 
The solid and dashed curves show the theoretical calculations
given by Cabrera et al.~\cite{CAB03} without and with
Pauli-blocking correction for the $\phi$ meson scattering angle
in the laboratory frame of 0${}^\circ$, respectively.
(b)~The ratio $P_{\rm out}/P_{\rm out}{\rm (Li)}$. 
The solid and dashed curves show the theoretical calculations same as (a).
}
\label{fig4}
\end{center}
\end{figure}
On the basis of the $\phi$ self-energy in the nuclear medium,
Cabrera {et al}.\@ presents
the $A$-dependence of the $\phi$ photo-production cross section from nuclei
in terms of the variable $P_{\rm out}\!=\!\sigma_A/(A\sigma_N)$,
which represents the probability of a photo-produced $\phi$ meson
going out a nucleus~\cite{CAB03}.
The $P_{\rm out}$ is deduced by using $Y$ described in Eq.~(\ref{eq:yld}) and
the normalized number of events for $\phi$ photo-production
on the nucleon,
$Y_0$, which is deduced from the present data
by using a Glauber-type multiple scattering theory.
Fig.~\ref{fig4}~(a) shows 
the obtained $P_{\rm out}$ in the experiment
and theoretical predictions given by Cabrera {et al}.\@
as a function of $A$.
It is noted that the averaged momentum of $\phi$ mesons in the present experiment is
$\langle P_\phi\rangle=1.8$~${\rm GeV}/c$,
while theoretical predictions are made for $P_\phi=2.0$~${\rm GeV}/c$.
The obtained $P_{\rm out}$ are smaller than the theoretical predictions.
The $\phi$ meson
flux obtained in the experiment is almost
half of the theoretical predictions.
The absolute value of $P_{\rm out}$ obtained in the experiment 
depends on an applied model to deduce $Y_0$.
However, the ratio of $P_{\rm out}/P_{\rm out}({\rm Li})$ is model-independent.
The ratios are smaller than the theoretical predictions as shown in Fig.~\ref{fig4}~(b).
The theoretical calculations underestimate the decrease of photo-produced
$\phi$ meson flux in the nucleus.
This discrepancy implies that 
the $\phi$-$N$ interaction is stronger than theoretical estimations due to the 
modification of the $\phi$ properties in the nuclear medium.

In summary, the photo-production of $\phi$ mesons from Li, C, Al, and Cu nuclei
at forward angles has been measured at $E_\gamma=1.5$--$2.4$~GeV.
The mass and width of the $\phi$ meson observed in the $K^+K^-$
invariant mass spectrum are consistent with those
of the free $\phi$ meson for all the nuclear targets used.
There is a possibility that 
the reconstructed invariant mass is insensitive to
possible in-medium
modification of the $\phi$ meson for the high momenta of the measured
$\phi$ mesons~({1.0--2.2~GeV$/c$}).

The $A$-dependence of the $\phi$ photo-production yields for
$|t|\le 0.6$~${\rm GeV}^2/c^2$ is found to be proportional to $A^{0.63\pm 0.05}$.
After subtracting the coherent contribution evaluated from the nuclear form factor,
the yields are found to have the $A$-dependence of $A^{0.72\pm 0.07}$.
The total cross section of the $\phi$-nucleon interaction, $\sigma_{\phi N}$,
is estimated to be $35^{+17}_{-11}$~mb from the $A$-dependence.
This value is much larger than $\sigma_{\phi N}$ in free space,
suggesting that the $\phi$ properties might change in the nuclear medium
although the change of the mass and width 
is not observed in the $K^+K^-$ invariant mass spectra.
The ratio $P_{\rm out}/P_{\rm out}({\rm Li})$ is smaller than the theoretical predictions,
which implies that the in-medium modification might be larger than the predictions.
\ack
The authors wish to thank the SPring-8 staff for their dedicated efforts
of providing a high quality beam.
They also thank all the people who contributed to
the construction of the LEPS facility.
They gratefully acknowledge Dr.~K.~Hasegawa (Honjo~Metal~Co.~Ltd.)
for making the Li target suitable for the present experiment.

\end{document}